\documentclass[conference]{IEEEtran}
\IEEEoverridecommandlockouts
\usepackage{cite}
\usepackage{amsmath,amssymb,amsfonts,yhmath}
\usepackage{algorithmic}
\usepackage{diagbox}
\usepackage{graphicx}
\usepackage{textcomp}
\usepackage{xcolor}
\usepackage{cuted}
\usepackage{subfig}

\def\BibTeX{{\rm B\kern-.05em{\sc i\kern-.025em b}\kern-.08em
    T\kern-.1667em\lower.7ex\hbox{E}\kern-.125emX}}
\begin{document}
\title{Radio Frequency Fingerprints Extraction for LTE-V2X:\! A\! Channel\! Estimation\! Based\! Methodology\
\thanks{© 2022 IEEE. Personal use of this material is permitted. Permission from IEEE must be obtained for all other uses, in any current or future media, including reprinting/republishing this material for advertising or promotional purposes, creating new collective works, for resale or redistribution to servers or lists, or reuse of any copyrighted component of this work in other works.}
}
\author{\IEEEauthorblockN{Tianshu Chen\IEEEauthorrefmark{1},
Hong Shen\IEEEauthorrefmark{1},
Aiqun Hu\IEEEauthorrefmark{1}\IEEEauthorrefmark{2},
Weihang He\IEEEauthorrefmark{3},
Jie Xu\IEEEauthorrefmark{3},
Hongxing Hu\IEEEauthorrefmark{4}}
\IEEEauthorblockA{\IEEEauthorrefmark{1}National Mobile Communications Research Laboratory, Southeast University, Nanjing, China}
\IEEEauthorblockA{\IEEEauthorrefmark{2}The Purple Mountain Laboratories for Network and Communication Security, Nanjing, China}
\IEEEauthorblockA{\IEEEauthorrefmark{3}School of Cyber Science and Engineering, Southeast University, Nanjing, China}
\IEEEauthorblockA{\IEEEauthorrefmark{4}China Automotive Innovation Corporation, Nanjing, China}
\IEEEauthorblockA{Email: \{iamtianshu, shhseu, aqhu, 220205165, 220205095\}@seu.edu.cn, huhongxing@t3caic.com}}

\maketitle

\begin{abstract}
The vehicular-to-everything (V2X) technology has recently drawn a number of attentions from both academic and industrial areas. However, the openness of the wireless communication system makes it more vulnerable to identity impersonation and information tampering. How to employ the powerful radio frequency fingerprint (RFF) identification technology in V2X systems turns out to be a vital and also challenging task. In this paper, we propose a novel RFF extraction method for Long Term Evolution-V2X (LTE-V2X) systems. In order to conquer the difficulty of extracting transmitter RFF in the presence of wireless channel and receiver noise, we first estimate the wireless channel which excludes the RFF. Then, we remove the impact of the wireless channel based on the channel estimate and obtain initial RFF features. Finally, we conduct RFF denoising to enhance the quality of the initial RFF. Simulation and experiment results both demonstrate that our proposed RFF extraction scheme achieves a high identification accuracy. Furthermore, the performance is also robust to the vehicle speed.
\end{abstract}

\begin{IEEEkeywords}
Vehicular-to-everything (V2X), radio frequency fingerprint (RFF), device identification, channel estimation, RFF denoising
\end{IEEEkeywords}

\section{Introduction}
\begin{figure*}[t]
\centerline{\includegraphics[width=18cm]{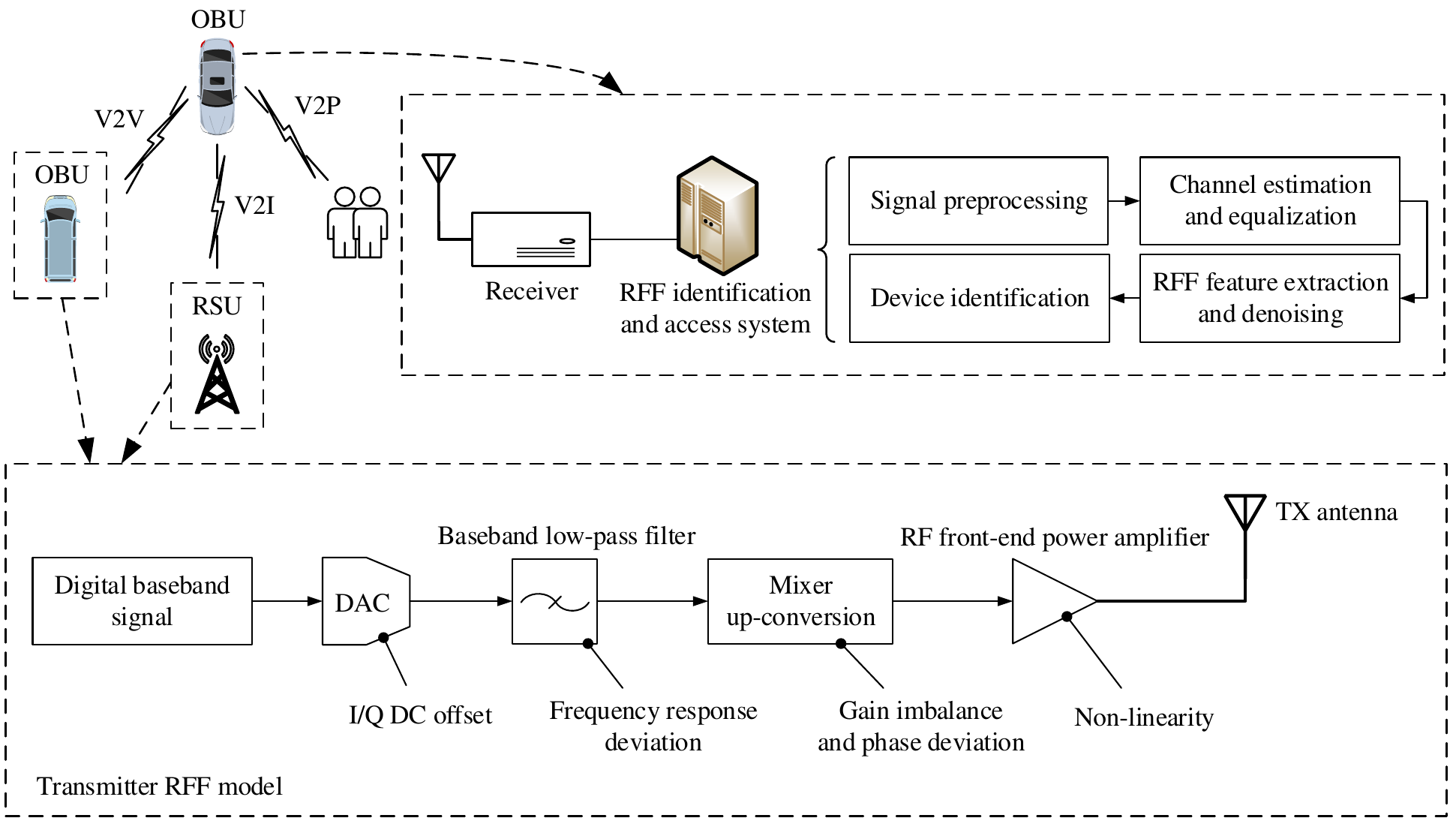}}
\caption{LTE-V2X RFF extraction and identification system framework and RFF model at the transmitter.}
\label{figmodel}
\end{figure*}
Vehicular-to-everything (V2X) has become a promising technique for intelligent transportation and autonomous driving. In particular, the cellular-V2X (C-V2X) has been widely acknowledged as a key V2X communication standard due to its superior performance \cite{9217500}, \cite{8891313}.

Since V2X relies on wireless transmission, the information is easy to be eavesdropped, forged or tampered with, which imposes great challenges on the safety of vehicles, pedestrians and road infrastructures in the V2X communication network \cite{9258954}. To deal with the security threats faced by wireless communications, there are usually two widely used authentication strategies: key-based cryptographic authentication and physical layer security-based non-cryptographic authentication \cite{9083673}. The cryptographic authentication technology needs to distribute and manage abundant communication keys, which occupies computing resources and leads to additional overhead and delays. Moreover, with the rapid development of computing capability of the computers, especially the emergence of quantum computers, traditional cryptography technologies are more vulnerable to brute-force attacks \cite{8490169}. On the contrary, the physical layer security based authentication has lower complexity and network overhead with lower latency compared to traditional cryptography-based authentication methods, and can achieve non-perceptual authentication without third-party facilities. One typical example is the radio frequency fingerprint (RFF) based authentication, which fully exploits the hardware differences between any two devices. Since the hardware characteristic of each device is unique and difficult to clone, the RFF based authentication can better resist the identity attacks and spoofing \cite{5601959}.

In literature, a variety of RFF extraction and identification methods have been advocated. Early works mainly focus on the characteristics of transient signals, such as instantaneous amplitude, frequency, and phase responses \cite{7960417}. Concerning the steady-state signal, such as preamble signals, researchers consider extracting the RFF features including I/Q offset\cite{5961627}, power spectral density \cite{4698196}, differential constellation trace figure \cite{8360937}. Furthermore, some universal RFF extraction methods which are independent of data, channel or modulation modes have also been studied. Concretely, Shen \textit{et al.} \cite{9715147} constructed channel independent spectrogram and utilized data augmentation for RFF extraction and identification of Lora devices, which achieves good performance under different channel conditions. Alternatively, Yang \textit{et al.} \cite{9519652} used random data segments to extract the tap coefficients of the least mean square (LMS) adaptive filter as data independent RFF. Sun \textit{et al.} \cite{9721428} verified the locality and inhomogeneity of the RFF distribution with the analysis in the cepstral domain, which yields modulation mode independent RFF.

The aforementioned works mainly consider the RFF extraction for low mobility and narrowband systems. However, for the V2X system, the channel typically varies fast due to the high mobility vehicles. In addition, the V2X signal usually has a large bandwidth which is more vulnerable to multipath environment. Therefore, the current RFF extraction methods for narrowband systems such as ZigBee and Lora cannot be directly applied for the V2X system because they do not take into account the impact of multipath and time-varying channels.

In this work, we propose a channel estimation based RFF extraction method for Long Term Evolution-V2X (LTE-V2X) systems, which, to the best of our knowledge, has not been investigated in existing works. Specifically, we first estimate the experienced wireless channel using an improved least square (LS) channel estimation method. Then, we perform channel equalization based on the channel estimate to obtain channel dependent RFF. The RFF quality is further enhanced via conducting time-domain denoising. It is worthwhile noting that the developed method eliminates the effect of the channel and the noise on the RFF with low implementation complexity, and can be extended to various broadband multi-carrier wireless communication systems.

This paper is organized as follows. Section II introduces the system model and signal preprocessing. Section III presents the details of the proposed RFF extraction methodology based on wireless channel estimation. Section IV evaluates the performance of the proposed RFF extraction method through simulations and experiments. Section V concludes this work.

\section{System Model and Signal Preprocessing}

\subsection{System Model}

Fig. \ref{figmodel} demonstrates the framework of the considered LTE-V2X RFF extraction and identification system together with the RFF model at the transmitter. More concretely, one V2X terminal, e.g., on board unit (OBU) or road side unit (RSU), first transmits data to other devices, where the transmitted signal includes the RFF of the transmitter. Then, the receiver preprocesses the received signal which consists of converting the RF signal to the baseband signal and performing time-frequency synchronization. Subsequently, the RFF features are extracted based on the synchronized signal, where the effects of the wireless channel and the noise on the RFF need to be mitigated. Finally, the device identification is performed using the extracted RFF features.

It is necessary to note that the considered RFF refers to all the characteristics of the circuits at the transmitter, which, as shown in Fig. 1, include the I/Q DC offsets of the digital-to-analog converter (DAC), the frequency response deviation of the filter, the gain imbalance and the carrier phase quadrature deviation of the mixer, and the non-linearity of the power amplifier \cite{Wang2016Wireless}.

\begin{figure}[t]
\centerline{\includegraphics[width=9cm]{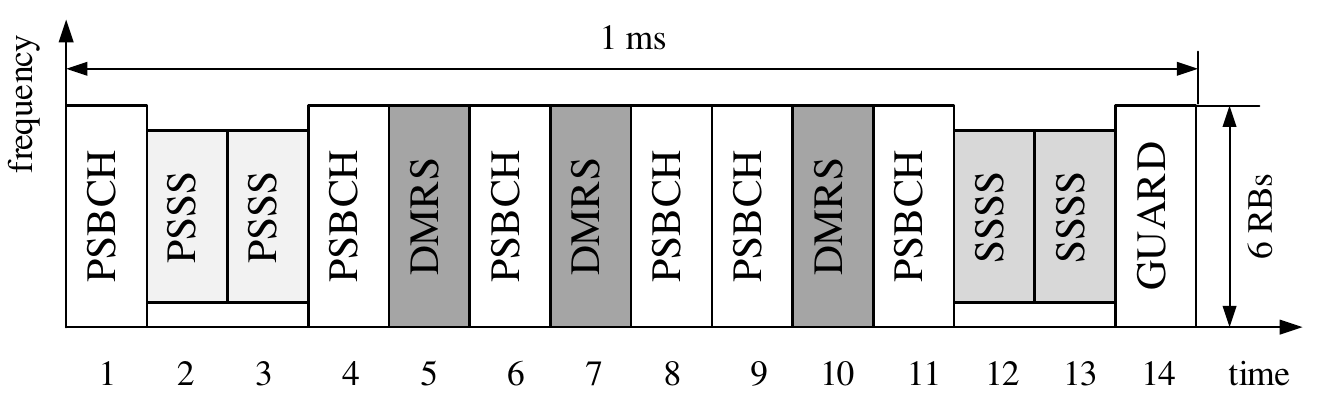}}
\caption{LTE-V2X PSBCH format.}
\label{fig1}
\end{figure}

\subsection{LTE-V2X PSBCH}

We adopt the physical sidelink broadcast channel (PSBCH) in LTE-V2X systems for RFF extraction. According to \cite{b5}, PSBCH is transmitted every 160 ms occupying the central 6 resource blocks (RBs), i.e., 72 subcarriers and 14 single-carrier frequency division multiple access (SC-FDMA) symbols.

The detailed format of PSBCH is shown in Fig. \ref{fig1}, where primary sidelink synchronization signal (PSSS), secondary sidelink synchronization signal (SSSS), and demodulation reference signal (DMRS) all depend on the currently used sidelink synchronization signal (SLSS) ID. Since the SLSS ID can be estimated \cite{b5}, we can readily obtain ideal PSSS, SSSS, and DMRS at the receiver which are used for extracting transmitter RFF.

\subsection{Signal Preprocessing}
In order to ensure the stability of the extracted RFF, the signal preprocessing procedure includes time synchronization and carrier frequency offset (CFO) estimation and compensation after the received signal is down-converted from the RF band to the baseband.

The time synchronization is realized by utilizing two identical training symbols, e.g., two repeated PSSS or SSSS symbols in LTE-V2X PSBCH, and the cross-correlation between the received signal $r(n)$ and the training signal $x(n)$ as
\begin{equation}
P(d)\!=\!\sum\limits_{n=0}^{N\!-\!1}\!\left|{r(n\!+\!d)x^*(n)}\right|^2\!+\!\sum\limits_{n=0}^{N\!-\!1}\!\left|{r(n\!+\!d\!+\!N\!+\!N_{C\!P})x^*(n)}\right|^2,
\label{eq21}
\end{equation}
where $N\!\!=\!\!2048$ for LTE-V2X systems and $N_{C\!P}$ denotes the length of the cyclic prefix (CP). When $P(d)$ exceeds a given threshold $P_{T\!H}$ and reaches the maximum, we obtain the estimated starting position of the training symbol \cite{1997Robust}, which is expressed by
\begin{equation}
\hat{d}=\arg\max_{d\in \{ d|P(d)>P_{T\!H}\}}P(d).
\label{eq22}
\end{equation}

Afterwards, the CFO is estimated by performing auto-correlation between adjacent two identical PSSS symbols and two identical SSSS symbols \cite{1997ML}, which is expressed as
\begin{equation}
\begin{aligned}
\hat{\varepsilon}=\frac{1}{2\pi(N\!+\!N_{C\!P})}\mathrm{angle}
\left\{ 
\sum\limits_{n=0}^{N-1}\!{[r(n\!+\!\hat d){r^*(n\!+\!\hat d\!+\!N\!+\!N_{C\!P})]}}\right. \\
\left. +\sum\limits_{n=0}^{N-1}\!{[r(n\!+\!\Delta n\!+\!\hat d){r^*(n\!+\!\Delta n\!+\!\hat d\!+\!N\!+\!N_{C\!P})]}}\right\},
\label{eq23}
\end{aligned}
\end{equation}
where $\mathrm{angle}\{\cdot \}$ returns the phase angle of the input complex number and $\Delta n$ represents the number of the sampling points between the first PSSS and the first SSSS. Accordingly, we obtain the CFO compensated signal by
\begin{equation}
y(n)=\tilde r(n){e}^{-j2\pi n\hat{\varepsilon}},
\label{eq24}
\end{equation}
where $\tilde r(n)$ denotes the time synchronized signal.

\section{Proposed RFF Extraction Method}
In this section, we propose a novel PSBCH based RFF extraction method for LTE-V2X systems, which mainly includes channel estimation, channel equalization, and RFF denoising.

\begin{figure}[t]
\begin{minipage}[b]{0.49\linewidth}
\centering
\subfloat[][initial time domain channel estimate $h_5(n)$]{\label{fig3_1}\includegraphics[width=1\linewidth]{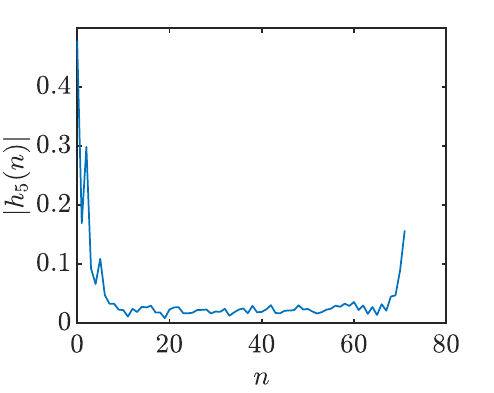}}
\end{minipage}
\medskip
\begin{minipage}[b]{0.49\linewidth}
\centering
\subfloat[][windowed time domain channel estimate $\hat h_5(n)$]{\label{fig3_2}\includegraphics[width=1\linewidth]{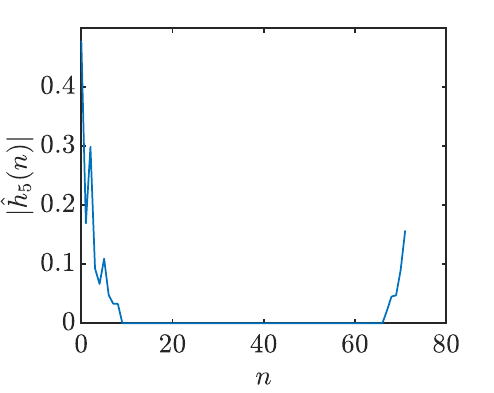}}
\end{minipage}

\caption{The initial and windowed time domain channel estimates of the DMRS symbol.}
\label{fig3}
\end{figure}

\subsection{Channel Estimation}
We adopt the improved LS algorithm \cite{Beek1995} for channel estimation. The main idea of the algorithm is to obtain the initial frequency domain channel estimate through the LS algorithm, which is then transformed into the time domain via inverse discrete Fourier transform (IDFT). Afterwards, we perform time-domain windowing to exclude the noise and the RFF. The resultant signal is finally transformed into the frequency domain via discrete Fourier transform (DFT). The detailed steps of channel estimation for the PSBCH subframe are described as follows.

Denote the $i$-th time-domain SC-FDMA symbol of the received PSBCH after preprocessing and CP removal by $y_i(n)$, which carries RFF information and channel information. Then, we transform the time-domain received signals corresponding to the PSSS, the SSSS, and the DMRS symbols into the frequency domain by performing DFT, which is expressed as
\begin{equation}
Y_i(k)=\text{DFT}_N\{{y_i(n)}\},0\le k\le N-1,
\label{eq1}
\end{equation}
where $\text{DFT}_N\{ \cdot \}$ denotes the $N$-point DFT and $i=2,3,5,7,$ $10,12,13$. Denote the frequency domain received signal corresponding to the effective bandwidth occupied by the PSSS, the SSSS, and the DMRS as $\wideparen Y_i(k)$. Then, the initial frequency domain channel estimate of the $i$-th symbol $\hat H_i(k)$ containing the RFF and the noise is calculated by
\begin{equation}
\hat H_i(k)=\frac{\wideparen Y_i(k)}{\wideparen X_i(k)}, k \in \mathbb N_i,
\label{eq2}
\end{equation}
where $\wideparen X_i(k)$ denotes the PSSS, the SSSS, or the DMRS, and $\mathbb N_i$ is defined by
\begin{equation}
\begin{aligned}
\mathbb N_i=\!\left\{ \begin{aligned}
&[5,66],i=2,3,12,13\\
&[0,71],i=5,7,10
\end{aligned} \right. .
\label{eq3}
\end{aligned}
\end{equation}

Subsequently, based on $\hat H_i(k)$, we obtain the initial time domain channel estimate by
\begin{equation}
\hat h_i(n)=\text{IDFT}_{N_i}\{{\hat H_i(k)}\}, n \in \mathbb N_i,
\label{eq4}
\end{equation}
where $\text{IDFT}_{N_i}\{{\cdot}\}$ denotes the $N_i$-point IDFT and $N_i$ is defined by
\begin{equation}
\begin{aligned}
N_i=\!\left\{ \begin{aligned}
&62,i=2,3,12,13\\
&72,i=5,7,10
\end{aligned} \right. .
\label{eq3_1}
\end{aligned}
\end{equation}

\begin{table*}[t]
\caption{RFF Parameters of 10 Simulated LTE-V2X Terminals}
\renewcommand\arraystretch{1.2}
\begin{center}

\begin{tabular}{c c c c c c}
\hline
Terminal index&DC offset&Filter coefficients&Gain imbalance&Phase deviation&Power amplifier coefficient\\
\hline
1&$D_I$=0, $D_Q$=0&$h_I$=[1 0], $h_Q$=[1 0]&0.1&0.1&[1 0 0] \\
2&$D_I$=0.01, $D_Q$=0&$h_I$=[1 0], $h_Q$=[1 0]&0.01&0.01&[1 0 0] \\
3&$D_I$=0, $D_Q$=-0.01&$h_I$=[1 0], $h_Q$=[1 0]&0&0&[1 0 0] \\
4&$D_I$=-0.005, $D_Q$=0.005&$h_I$=[1 0], $h_Q$=[1 0]&0.01&0.01&[1 0 0] \\
5&$D_I$=0.005, $D_Q$=-0.005&$h_I$=[1 0], $h_Q$=[1 0]&0&0&[1 0 0] \\
6&$D_I$=0, $D_Q$=0&$h_I$=[1 0], $h_Q$=[1 0]&0.05&0&[0.9+0.15j 0.1 0.1-0.15j] \\
7&$D_I$=0, $D_Q$=0&$h_I$=[1 0], $h_Q$=[1 0]&0&0.05&[1.15 -0.2 0] \\
8&$D_I$=0, $D_Q$=0&$h_I$=[0.825 0], $h_Q$=[1.175 0]&0&0&[1 0 0] \\
9&$D_I$=0, $D_Q$=0&$h_I$=[1 0.175], $h_Q$=[1 -0.175]&0&0&[1 0 0] \\
10&$D_I$=0.005, $D_Q$=0&$h_I$=[0.95 0], $h_Q$=[1 0.05]&0.05&0.05&[0.95-0.05j 0 0] \\
\hline
\end{tabular}
\label{simulate}
\end{center}
\end{table*}

Since the channel impulse response is concentrated in a few time domain samples while the noise and the RFF are distributed over the entire time domain, we can apply an appropriate window on $\hat h_i(n)$ to obtain an improved time domain channel estimate by
\begin{equation}
\breve{h}_i(n)=\hat h_i(n)w_i(n), n \in \mathbb N_i,
\label{eq5}
\end{equation}
where $w_i(n)$ denotes the window function. Fig. \ref{fig3} illustrates the windowing operation, where a rectangular window is used. Since most noises and RFFs are removed by the windowing operation, the resultant channel estimate becomes more accurate.

After obtaining $\breve{h}_i(n)$, we further acquire the corresponding frequency domain channel estimate as
\begin{equation}
\breve{H}_i(k)=\text{DFT}_{N_i}\{{\breve h_i(n)}\},k \in \mathbb N_i,
\label{eq8}
\end{equation}
Considering the fact that the channels experienced by adjacent symbols are approximately identical, especially when the vehicle speed is not very high, we can further average adjacent $\breve{H}_i(k)$'s to suppress the noise, thus improving the channel estimation accuracy. For instance, if the channel variation in one subframe is negligible, the ultimate frequency domain channel estimate can be calculated by
\begin{equation}
\tilde{H}(k)\!=\!\left\{ \begin{aligned}
  & \frac{{{{\breve{H}}}_{P\!S\!S\!S}}(k)+{{{\breve{H}}}_{D\!M\!R\!S}}(k)+{{{\breve{H}}}_{S\!S\!S\!S}}(k)}{7},5\le k\le 66 \\ 
 & \frac{{{{\breve{H}}}_{D\!M\!R\!S}}(k)}{3},0\le k\le 71 \\ 
\end{aligned} \right. ,
\label{eq9}
\end{equation}
where
\begin{equation}
{{{\breve{H}}}_{P\!S\!S\!S}}(k)={{{\breve{H}}}_{2}}(k)+{{{\breve{H}}}_{3}}(k),
\label{eq9_1}
\end{equation}
\begin{equation}
{{{\breve{H}}}_{D\!M\!R\!S}}(k)={{{\breve{H}}}_{5}}(k)+{{{\breve{H}}}_{7}}(k)+{{{\breve{H}}}_{10}}(k),
\label{eq9_2}
\end{equation}
\begin{equation}
{{{\breve{H}}}_{S\!S\!S\!S}}(k)={{{\breve{H}}}_{12}}(k)+{{{\breve{H}}}_{13}}(k).
\label{eq9_3}
\end{equation}

\subsection{Channel Equalization}
After acquiring the channel estimate $\tilde H(k)$, we can perform channel equalization to remove the channel information and achieve the initial RFF features $R_i(k)$ by
\begin{equation}
R_i(k)=\frac{\wideparen Y_i(k)}{\tilde{H}(k)}, k \in \mathbb N_i.
\label{eq7}
\end{equation}
Note that the above channel equalization will not lead to a loss of RFF information since most RFFs have been removed by the windowing operation during the channel estimation stage.

\subsection{RFF Denoising}
According to \eqref{eq7}, the initial RFF feature is still affected by the noise in $\wideparen Y_i(k)$. To alleviate the impact of noise on the extracted RFF, we further average the initial RFFs corresponding to the same data sequence. Specifically, the denoised RFFs for the PSSS, the DMRS, and the SSSS are given by
\begin{equation}
R_{P\!S\!S\!S}(k)=\frac{R_2(k)+R_3(k)}{2},5\le k\le 66,
\label{eq10}
\end{equation}
\begin{equation}
\begin{aligned}
R_{D\!M\!R\!S}(k)\!=\!\left\{ \begin{aligned}
&\frac{ {{R}_{5}}(k)+{{R}_{7}}(k)+{{R}_{10}}(k) }{3},N_{I\!D}^{S\!L}\bmod 2\!=\!0\\
&\frac{{{R}_{5}}(k)+{{R}_{10}}(k) }{2},\ \ \ \ \ \ \ \ \ \ \ N_{I\!D}^{S\!L}\bmod 2\!=\!1
\end{aligned} \right. ,\\0\le k\le 71,
\label{eq11}
\end{aligned}
\end{equation}
\begin{equation}
R_{S\!S\!S\!S}(k)=\frac{R_{12}(k)+R_{13}(k)}{2},5\le k\le 66.
\label{eq12}
\end{equation}
Note that the DMRS sequence on the 7th symbol differs from those on the 5th and 10th symbols when the SLSS ID $N_{I\!D}^{S\!L}$ is odd. Hence, for this case, we only calculate the mean of ${{R}_{5}}(k)$ and ${{R}_{10}}(k)$ which have the same data sequence. Finally, we obtain ultimate RFF features $\mathbf R(k)$ as
\begin{equation}
\begin{aligned}
\mathbf R(k)=\left\{ \begin{aligned}
 R_{D\!M\!R\!S}(k),\ \ \ \ \ \ \ \ \ \ \ \ 0\le k\le 4,67\le k\le 71\\
\left[ R_{P\!S\!S\!S}(k),R_{D\!M\!R\!S}(k),R_{S\!S\!S\!S}(k) \right],5\le k\le 66
\end{aligned} \right. .
\label{eq13}
\end{aligned}
\end{equation}

\section{Simulation and Experiment Results}
In the experiment, we employ 10 simulated LTE-V2X terminals with different RFF parameters and 6 actual LTE-V2X modules to generate PSBCH subframes, respectively, and evaluate the classification performance of different devices based on our proposed RFF extraction scheme.

\subsection{Simulation Verification}
For the simulation, we set different RFF parameters for 10 terminals, including the I/Q DC offsets, the baseband low-pass filter coefficients, the gain imbalance, the phase quadrature deviation, and the RF front-end power amplifier coefficients, which are specifically shown in Table \ref{simulate}, to ensure the modulation domain error vector magnitude (EVM) is within 17.5\% \cite{std36101}.

Next, the PSBCH signals carrying the RFFs generated by the 10 terminals pass through the simulated extended typical urban (ETU) multipath channel \cite{std36104}, where the vehicle speed ranges from 0 to 120 km/h. Moreover, the SNR ranges from 0 to 30 dB.

\begin{figure}[t]
\centerline{\includegraphics[width=9cm]{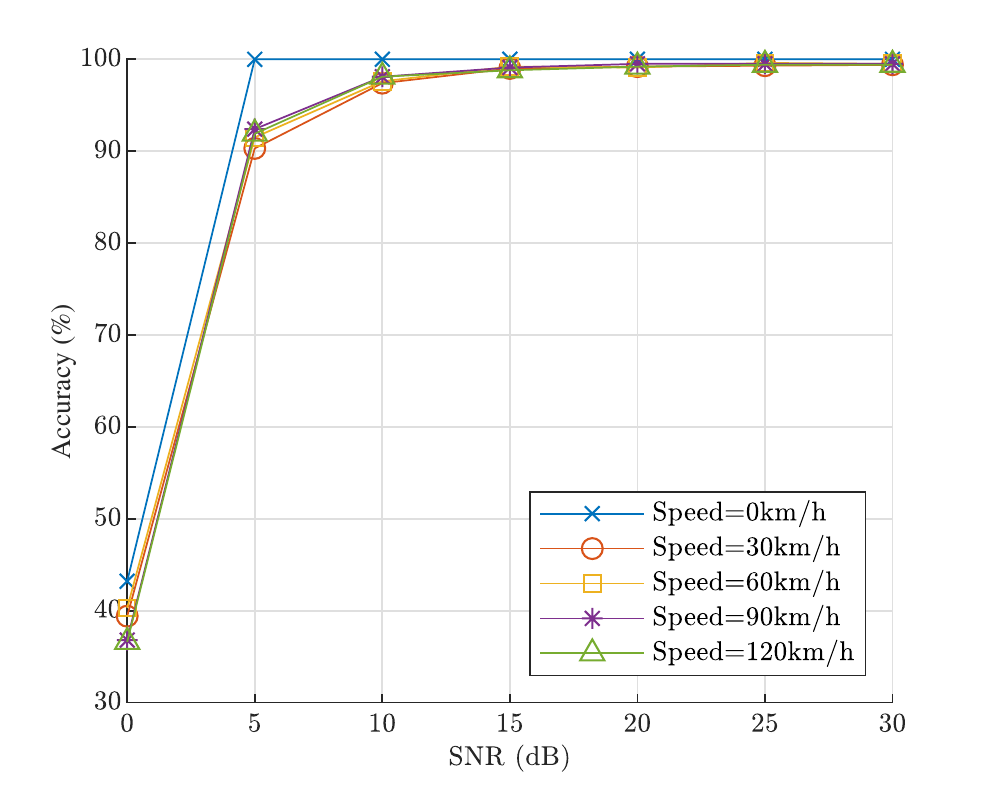}}
\caption{Identification accuracy of 10 simulated LTE-V2X terminals based on the proposed RFF extraction method under different SNRs and different vehicle speeds.}
\label{acc1}
\end{figure}

\begin{figure}[t]
\centerline{\includegraphics[width=9cm]{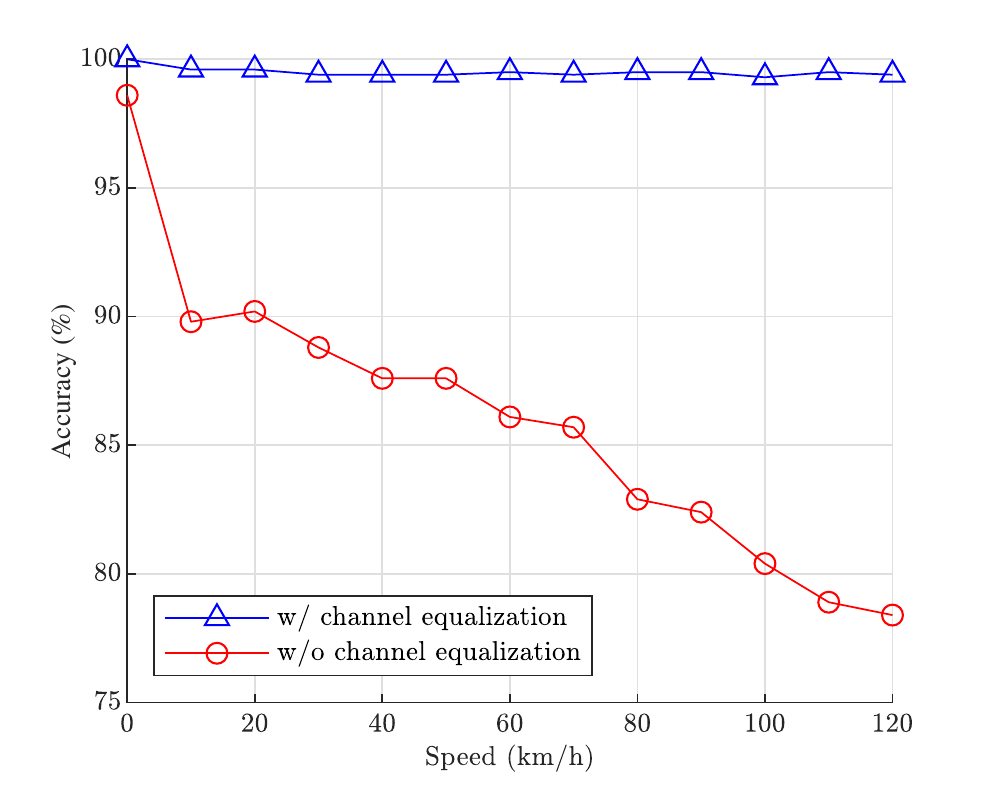}}
\caption{Comparison of the identification accuracy of 10 simulated LTE-V2X terminals with and without channel equalization (SNR = 30 dB).}
\label{acc2}
\end{figure}

Then, we conduct classification experiments on 10 terminals using random forest algorithm. The 700 received PSBCH subframes of each terminal constitute the training set, where the SNR is 30 dB and the vehicle speed is 30 km/h. The test set consists of 300 other subframes from each terminal. The identification accuracy of the 10 terminals under different SNRs and different vehicle speeds is depicted in Fig. \ref{acc1}. It can be found that the vehicle speed has little effect on the RFF identification accuracy rate. When the SNR exceeds 10 dB, the accuracy always remains above 97\% regardless of the speed, while the accuracy decreases significantly when the SNR drops below 10 dB mainly because we only use one PSBCH subframe for RFF extraction. It reveals that the proposed RFF extraction method has excellent classification performance under medium and high SNRs.

Fig. \ref{acc2} compares the RFF identification performances of the methods with and without channel equalization, where the SNR is 30 dB. When the speed increases from 0 to 120 km/h, there is no obvious loss in the accuracy rate for the channel equalization based method, which always remains over 99\%, while the identification accuracy without channel equalization falls rapidly especially at high speeds, which indicates that our proposed method based on channel estimation can effectively mitigate the impact of wireless channels on the RFF extraction. 

\subsection{Experiment Verification}

\begin{figure}[t]
\centerline{\includegraphics[width=9cm]{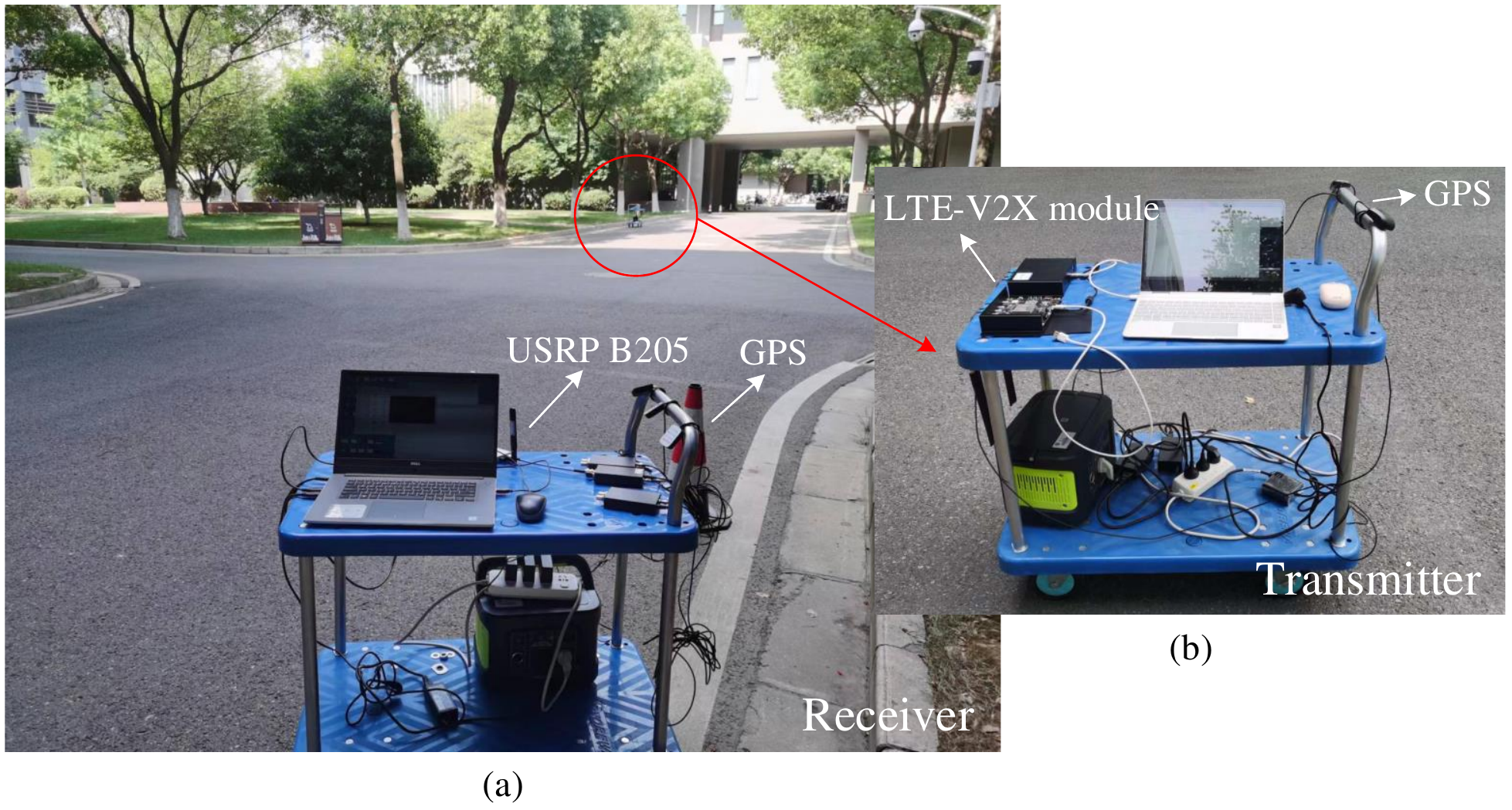}}
\caption{Experiment setup: (a) receiving device (USRP B205); (b) transmitting device (LTE-V2X module).}
\label{lab}
\end{figure}

\begin{table}[t]
\caption{RFF Identification Accuracy of 6 LTE-V2X Modules Under Different Speeds}
\renewcommand\arraystretch{1.2}
\begin{center}

\setlength{\tabcolsep}{2mm}
\begin{tabular}{c c c c c}
\hline
\diagbox[dir=SE]{Device}{Accuracy}{Speed}&0 km/h&10 km/h&20 km/h&30 km/h\\
\hline
Module 1&92\%&93\%&90\%&91\%\\
Module 2&69\%&71\%&69\%&68\%\\
Module 3&92\%&90\%&93\%&93\%\\
Module 4&100\%&100\%&100\%&97\%\\
Module 5&100\%&100\%&100\%&100\%\\
Module 6&100\%&100\%&100\%&100\%\\
Average&92.2\%&92.3\%&92\%&91.5\%\\
\hline
\end{tabular}
\label{experiment}
\end{center}
\end{table}

For the experiment, we use 6 LTE-V2X modules to transmit PSBCH subframes and utilize USRP B205 to receive the signals. The experiment setup is shown in Fig. \ref{lab}. First, we collect 400 PSBCH subframes for each module as training set under static state and low-speed moving state. Subsequently, 100 other subframes are captured from each module as test set, where the speed ranges from 10 to 30 km/h. The classification accuracy of the 6 LTE-V2X modules are shown in Table \ref{experiment}. It can be seen that the average accuracy exceeds 90\%. Moreover, the accuracy rate does not drop significantly after the speed increases. Note that modules 1 to 4 belong to the same type with very similar RFF features. Hence, the corresponding classification accuracy is relatively low.
\section{Conclusion}
In this paper, we proposed a novel RFF extraction method for LTE-V2X systems. Focusing on the PSSS, the SSSS, and the DMRS of PSBCH, we successfully obtained highly distinguishable RFF features by performing channel estimation, channel equalization, and RFF denoising. As verified via both simulations and experiments, our method displays robust performance under challenging time-varying and multipath channels. The proposed method can also be applied to any broadband multi-carrier communication systems that have fixed sequences. In the future work, more terminals can be tested in practical high mobility channel environments to further verify the effectiveness of this method.

\end{document}